\documentclass{jpsj-suppl}
\usepackage{txfonts} 

\title{Excited Scalar Mesons and the Search for Glueballs}

\author{Denis \textsc{Parganlija}$^{1}$ and Francesco \textsc{Giacosa}$^{2}$}

\inst{$^{1}$
Institut f\"{u}r Theoretische Physik, Technische Universit\"{a}t Wien, Wiedner Hauptstr.\ 8-10, 1040 Vienna, Austria
\\
$^{2}$ Institute of Physics, Jan Kochanowski University, 25406 Kielce, Poland}

\email{denisp@hep.itp.tuwien.ac.at}

\recdate{June 15, 2016}

\abst{We discuss the importance of mixing between glueballs -- bound states of gluons --
and excited $\bar{q}q$ states for the glueball search. A preliminary study of the excited states in the Extended Linear Sigma
Model (eLSM) suggests their masses in the scalar channel to be in the vicinity of the scalar-glueball mass found in 
Lattice QCD. This could have implications for future glueball searches.
\\
This article has been prepared for Proceedings of the 12th International Conference on Low Energy Antiproton Physics, 
Kanazawa/Japan, March 6-11, 2016.}

\kword{meson, glueball, quarkonium}

\begin{document}
\maketitle

\section{Meson Spectrum of Quantum Chromodynamics}

The low-energy region of Quantum Chromodynamics (QCD) -- the theory of strong interaction -- exhibits two interesting phenomena:
large value of the strong coupling \cite{AF} and confinement (see, e.g., Ref.\ \cite{Alkofer}). A consequence is the emergence
of an abundant spectrum of hadrons -- states composed of QCD degrees of freedom, quarks and gluons. In addition to hadrons
with half-integer spin (baryons), a vast class of integer-spin states also appears in the experimental data;
they are denoted as mesons. Listings of
the Particle Data Group (PDG) contain several hundreds of mesons \cite{PDG} but even after decades of research the internal
structure in terms of quarks and gluons has still not been clarified for many of these states, particularly for those with
masses below $\lesssim 2$ GeV.
\\
\\
On the theory side, there are numerous possibilities
to combine (constituent) quarks and gluons into mesons: as $\bar{q}q$ states (quarkonia), $\bar{q}\bar{q}qq$ states 
(tetraquarks), glueballs (bound states of gluons) and others \cite{Jaffe}. States of different structure should possess distinct observable features, most notably masses and decay patterns, allowing
for their identification in experimental data. The problem is that physical mesons emerge from
mixing of pure states that have the same quantum numbers. Separating the individual contributions to such mixed states is then a non-trivial
task, both for theory and experiment.
\\
\\
A prominent example is the channel of scalar isosinglet mesons, defined as having $IJ^{PC} = 00^{++}$ where $I$, $J$, $P$ and $C$
are respectively the quantum numbers of isospin, total spin, parity and charge conjugation. Six of these so-called $f_0$ states are listed
by the PDG in the energy region up to 2 GeV: $f_0(500)$, $f_0(980)$, $f_0(1370)$, $f_0(1500)$, $f_0(1710)$ and 
$f_0(2020)$ \cite{PDG}; the last one is currently unconfirmed. Additionally, a seventh state -- the $f_0(1790)$ resonance --
was found by the LHCb \cite{LHCb} and, prior to that, BES \cite{BES} Collaborations. The $f_0$ spectrum is then clearly oversaturated
if viewed only in terms of one class of pure states. 
\\
\\
Mixing of $\bar{n}n$ and $\bar{s}s$ quarkonia ($n$: up and down quarks) with the scalar glueball has been proposed as an explanation
for the emergence of at least a part of the $f_0$ spectrum. The $f_0(1500)$ and $f_0(1710)$ resonances have been alternatingly 
discussed as states with a large glueball component \cite{glueballmixing,Janowski:2011gt}; a recent approach from large-$N_c$ holographic
QCD has even suggested $f_0(1710)$ to be compatible with a pure glueball \cite{BPR}.
\\
\\
The mentioned mixing scenarios are based on the assumption that the scalar glueball mixes only with 
ground-state $\bar{q}q$ mesons. This is, however,
not the only option: mixing may also occur with the first \emph{excitations} of $\bar{q}q$ states. Glueball properties may change in this case -- and this is of high
importance for experimental glueball searches \cite{PANDA, NICA}.
However, the excited $\bar{q}q$ states can only influence other mesons if the masses 
are reasonably close to each other. Thus, in this article, a preliminary study 
of excited-quarkonia masses is presented; further phenomenology of these
states will be discussed in Ref.\ \cite{PG}.

\section{Excited Quarkonia and eLSM}

\begin{figure}[h]
\begin{center}%
\begin{tabular}
[c]{cc}%
\resizebox{78mm}{!}{\includegraphics{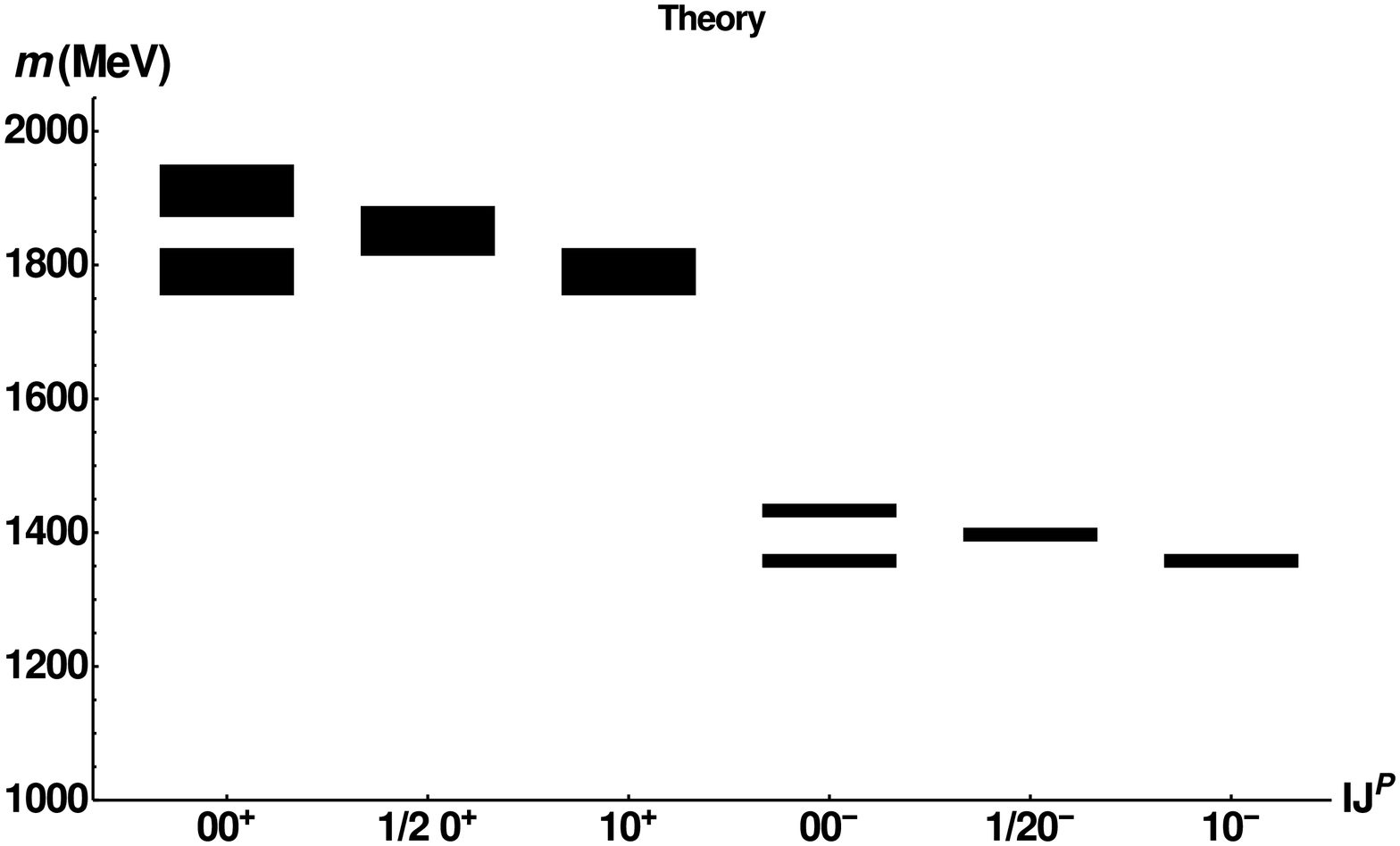}} &
\resizebox{78mm}{!}{\includegraphics{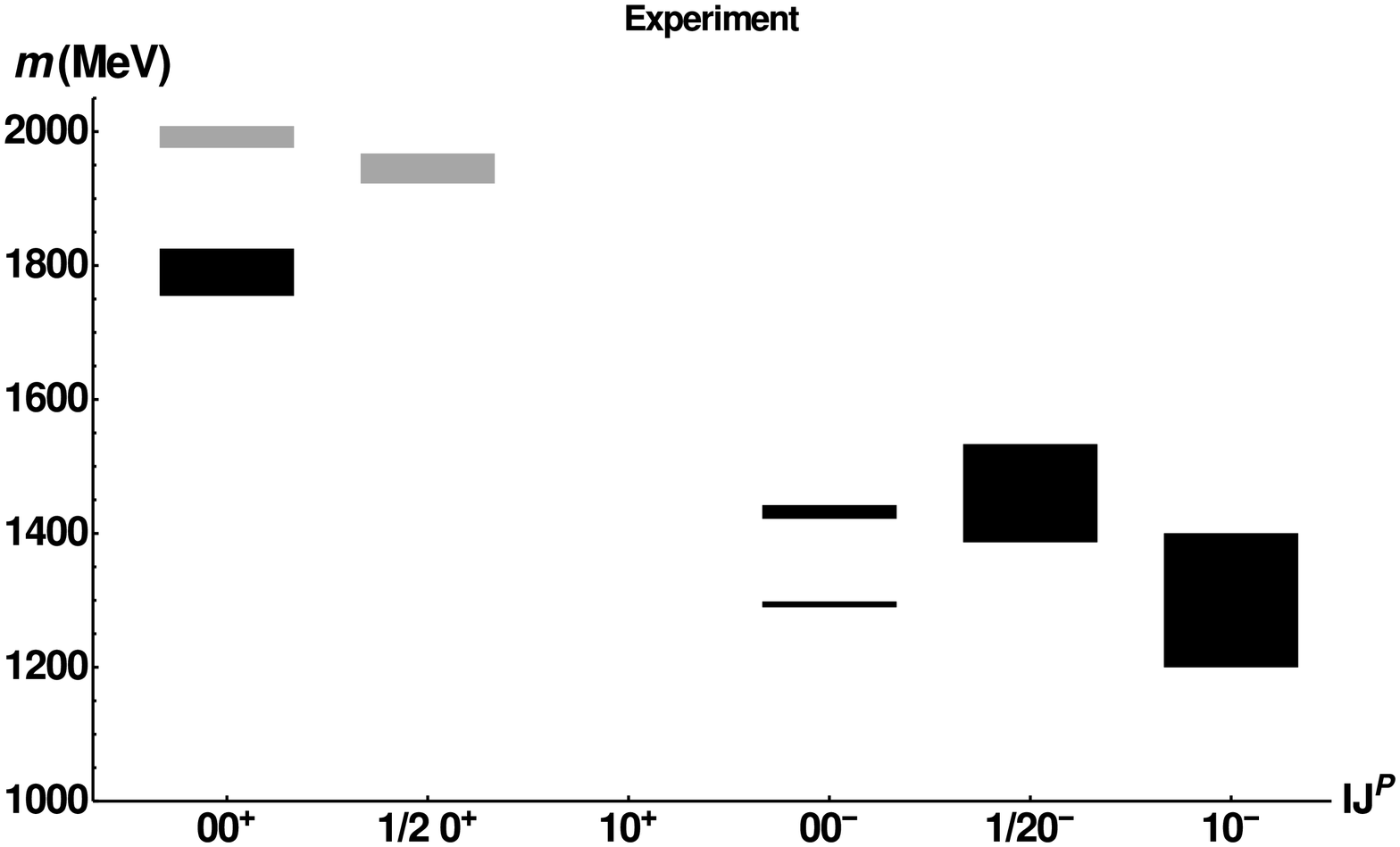}}
\end{tabular}
\end{center}
\caption{Preliminary masses of unmixed/pure excited $\bar{q}q$ states from the Extended Linear Sigma Model (left) and masses from the experimental
data (right). Area thickness corresponds to mass uncertainties on both panels. The 
lower $00^+ (\equiv \sigma_N^E)$, higher $00^-(\equiv \eta_S^E)$ as well as $1/20 ^- (\equiv K^E)$ 
and $10^- (\equiv \pi^E)$ states from the left panel were used as input. 
There is a remarkable proximity of the predicted masses
for the $1/20^+ (\equiv K_{0}^{\star E})$ and the upper $00^+ (\equiv \sigma_S^E)$ states to the $K_0^\star (1950)$ and $f_0(2020)$
resonances (that are lightly shaded on the right since they are denoted by the PDG as unconfirmed \cite{PDG}). The 
$10^+ (\equiv a_0^E)$ state does not (yet) have an experimentally measured counterpart.}
\label{Figure1}%
\end{figure}

Excited mesons were first studied several decades ago \cite{Freund:1968nf}; to date, they have been considered in a wide range of 
approaches including QCD models / chiral Lagrangians \cite{Ino:1984pi}, Lattice QCD \cite{Lacock:1996vy}, Bethe-Salpeter equation 
\cite{Holl:2004fr}, NJL Model and its extensions \cite{Volkov:1997dd}, light-cone models \cite{Arndt:1999wx}, QCD string approaches
\cite{Badalian:2002rc} and QCD domain walls \cite{Nedelko:2016gdk}. Chiral symmetry has also been suggested to become effectively
restored in excited mesons \cite{Wagenbrunn:2006cs}.
\\
\\
In this article, excited mesons in the scalar and pseudoscalar channels are studied by means of an Extended Linear Sigma Model
(eLSM), a wide-ranging approach to non-perturbative QCD incorporating (\textit{i}) chiral symmetry and its breaking; (\textit{ii})
dilatation symmetry and its breaking; (\textit{iii}) symmetry under charge conjugation, parity and time reversal, respectively;
(\textit{iv}) ground-state scalar, pseudoscalar, vector and axial-vector $\bar{q}q$ states in three
flavours; (\textit{v}) a pure scalar dilaton (glueball) state and (\textit{vi}) the first excitations in the three-flavour scalar and
pseudoscalar channels. The model without the excited states has already been 
used to study broad phenomeonology of mesons, see Refs.\ \cite{Janowski:2011gt,PGR,PKWGR}
and references therein.\\
\\
The eLSM Lagrangian is constructed analogously to that presented in Ref.\ \cite{Janowski:2011gt}; as indicated above, only masses of the
excited quarkonia will be discussed in this article for which the relevant part of the Lagrangian reads
\begin{align}
\mathcal{L}_{\mbox{\tiny mass}} &
 = (m_{0}^{\ast})^{2}\left(  \frac{G}{G_{0}}\right)  ^{2}\mathop{\mathrm{Tr}}(\Phi
_{E}^{\dagger}\Phi_{E})
-\lambda_{1}^{\ast}\mathop{\mathrm{Tr}} (\Phi_{E}^{\dagger}\Phi
_{E} )\mathop{\mathrm{Tr}}(\Phi^{\dagger}\Phi) -\lambda_{2}^{\ast}\mathop{\mathrm{Tr}}(\Phi_{E}^{\dagger}\Phi_{E}%
\Phi^{\dagger}\Phi+\Phi_{E}\Phi_{E}^{\dagger}\Phi\Phi^{\dagger}) \nonumber\\
& -\kappa_{2}%
[\mathop{\mathrm{Tr}}(\Phi_{E}^{\dagger}\Phi+\Phi^{\dagger}\Phi_{E})]^{2}%
-\xi_{2}\mathop{\mathrm{Tr}}(\Phi
_{E}^{\dagger}\Phi\Phi_{E}^{\dagger}\Phi+\Phi^{\dagger}\Phi_{E}\Phi^{\dagger
}\Phi_{E}) + \mathop{\mathrm{Tr}}(\Phi_{E}^{\dagger}\Phi_{E}\delta_E+\Phi_{E}\Phi
_{E}^{\dagger}\delta_E)\text{ .}%
\label{Lagrangian}%
\end{align}
$\,$ \\
In Eq.\ (\ref{Lagrangian}), $G$ is a scalar dilaton field whose Lagrangian describes the trace anomaly of QCD \cite{Rosenzweig:1979ay}
and $G_0$ is the expectation value of $G$; here, $G = G_0$ since only quarkonia will be discussed below.
$\Phi$ is the multiplet of scalar $S$ and pseudoscalar $P$ quarkonia reading [$T_i$: generators of $U(N_f=3)$]:

\begin{align}
\Phi  &  =\sum_{i=0}^{8}(S_{i}+iP_{i})T_{i}= \frac{1}{\sqrt{2}}\left(
\begin{array}
[c]{ccc}%
\frac{(\sigma_{N}+a_{0}^{0})+i(\eta_{N}+\pi^{0})}{\sqrt{2}} & a_{0}^{+}%
+i\pi^{+} & K_{0}^{\star+}+iK^{+}\\
a_{0}^{-}+i\pi^{-} & \frac{(\sigma_{N}-a_{0}^{0})+i(\eta_{N}-\pi^{0})}%
{\sqrt{2}} & K_{0}^{\star0}+iK^{0}\\
K_{0}^{\star-}+iK^{-} & {\bar{K}_{0}^{\star0}}+i{\bar{K}^{0}} & \sigma
_{S}+i\eta_{S}%
\end{array}
\right)  \text{ .}\label{Phi}
\end{align}
$\,$\\\\
$\Phi_E$, the multiplet of excited scalar and pseudoscalar quarkonia, is then defined analogously to that of Eq.\ (\ref{Phi}):
\begin{align}
\Phi_E  & = \frac{1}{\sqrt{2}}\left(
\begin{array}
[c]{ccc}%
\frac{(\sigma_{N}^E+a_{0}^{0E})+i(\eta_{N}^E+\pi^{0E})}{\sqrt{2}} & a_{0}^{+E}%
+i\pi^{+E} & K_{0}^{\star+ E}+iK^{+E}\\
a_{0}^{-E}+i\pi^{-E} & \frac{(\sigma_{N}^E-a_{0}^{0E})+i(\eta_{N}^E-\pi^{0E})}%
{\sqrt{2}} & K_{0}^{\star0 E}+iK^{0 E}\\
K_{0}^{\star- E}+iK^{-E} & {\bar{K}_{0}^{\star0 E}}+i{\bar{K}^{0 E}} & \sigma
_{S}^E + i\eta_{S}^E%
\end{array}
\right)  \text{ .}\label{PhiE}
\end{align}
$\,$\\
The Lagrangian in Eq.\ (\ref{Lagrangian}) is invariant under global chiral $U(3)_L \times U(3)_R$ transformations since $\Phi$ and
$\Phi_E$ transform as $\Phi_{(E)} \rightarrow U_L \Phi_{(E)} U_R^{\dagger}$. Spontaneous breaking of the chiral symmetry is implemented
by introducing non-vanishing vacuum expectation values $\phi_N$ and $\phi_S$ of the non-strange and strange scalar isosinglets 
$\sigma_N$ and $\sigma_S$, respectively
(but not for their excited counterparts \cite{Giacosa:2009qq}). Explicit chiral-symmetry breaking is implemented by the constant matrix
$\delta_E = \mbox{diag}(0,0,\delta)$ where the first two entries are proportional to masses of up and down quarks ($\simeq$ 0) and the
third entry to the $s$-quark mass.
\\
\\
The mass terms of the excited quarkonia are as follows:

\begin{align}
m_{\sigma_N^E}^2 & = ( m_{0}^{\ast} )^{2} + \left ( 2 \kappa_2 + \frac{\lambda_{1}^{\ast} + \lambda_{2}^{\ast} + \xi_2}{2} \right )\phi_N^2 
+ \frac{\lambda_{1}^{\ast}}{2} \phi_S^2 \label{sigmaNE} \\
m_{\pi^E}^2 & = m_{\eta_N^E}^2 = ( m_{0}^{\ast} )^{2} +\frac{\lambda_{1}^{\ast} + \lambda_{2}^{\ast} - \xi_2}{2} \phi_N^2 
+ \frac{\lambda_{1}^{\ast}}{2} \phi_S^2 \\
m_{a_0^E}^2 & = ( m_{0}^{\ast} )^{2} + \frac{\lambda_{1}^{\ast} + \lambda_{2}^{\ast} + \xi_2}{2} \phi_N^2 
+ \frac{\lambda_{1}^{\ast}}{2} \phi_S^2 \\
m_{\sigma_S^E}^2 & = ( m_{0}^{\ast} )^{2} - 2 \delta + \frac{\lambda_{1}^{\ast}}{2} \phi_N^2 
+ \left ( 2 \kappa_2 + \frac{\lambda_{1}^{\ast}}{2} + \lambda_{2}^{\ast} + \xi_2 \right )\phi_S^2 \\
m_{\eta_S^E}^2 & = ( m_{0}^{\ast} )^{2} - 2 \delta + \frac{\lambda_{1}^{\ast}}{2} \phi_N^2
+ \left ( \frac{\lambda_{1}^{\ast}}{2} + \lambda_{2}^{\ast} - \xi_2 \right )\phi_S^2  \\
m_{K_{0}^{\star \, E}}^2 & =  ( m_{0}^{\ast} )^{2} - \delta 
+ \left( \frac{\lambda_{1}^{\ast}}{2} + \frac{\lambda_{2}^{\ast}}{4} \right ) \phi_N^2
+ \frac{\xi_2}{\sqrt{2}} \phi_N \phi_S +  \frac{\lambda_{1}^{\ast} + \lambda_{2}^{\ast}}{2} \phi_S^2 \\
m_{K^{ E}}^2 & =  ( m_{0}^{\ast} )^{2} - \delta 
+ \left( \frac{\lambda_{1}^{\ast}}{2} + \frac{\lambda_{2}^{\ast}}{4} \right ) \phi_N^2
- \frac{\xi_2}{\sqrt{2}} \phi_N \phi_S +  \frac{\lambda_{1}^{\ast} + \lambda_{2}^{\ast}}{2} \phi_S^2 \label{KE}
\end{align}
$\,$\\
Six parameters are present in the mass terms; two of them, $\lambda_{1}^{\ast}$ and $\kappa_2$, are suppressed
in the limit of large number of colours and are therefore, as a first approximation, set to zero. The remaining parameters
($m_{0}^{\ast}$, $\lambda_{2}^{\ast}$, $\xi_2$ and $\delta$)
must then be
determined from at least four experimentally known meson masses. (Note that the values of the condensates $\phi_N$ and $\phi_S$
were determined in Ref.\ \cite{PKWGR}.) Previous studies relying on the model used in this article have
found ground-state scalar quarkonia to populate the energy region above 1 GeV \cite{PGR,PKWGR}; the excited scalar $\bar{q}q$ states would then
need to emerge as multiplets with exactly the same quantum numbers and even higher masses. However, the availability of experimental candidates for excited scalars -- as well as pseudoscalars --
is at times limited or their identification is unclear:

\begin{itemize}
 \item In the scalar ($J^P = 0^+$) channel, resonances with masses between 1 GeV and 2 GeV listed by the PDG \cite{PDG} are $f_0(1370)$, $f_0(1500)$, $f_0(1710)$, $a_0(1450)$, 
 $K_0^\star(1430)$, $K_0^\star (1950)$ and $f_0(2020)$. The first five are assumed to represent ground-state quarkonia with glueball admixture
 in the $f_0$ resonances (see Refs.\ \cite{Janowski:2011gt,PGR,PKWGR} and references therein);
the last two currently lack confirmation. In this article, the $f_0(1790)$ resonance \cite{BES,LHCb} is assumed to represent
the excited scalar isosinglet $\bar{n}n$ state due to its predominant decay into
pions while the kaon decay is strongly suppressed. Consequently it is assigned to $\sigma_N^E$. 

\item In the pseudoscalar ($J^P = 0^-$) channel, resonances with masses between 1 GeV and 2 GeV listed by the PDG \cite{PDG} are
$\eta(1295)$, $\pi(1300)$, $\eta(1405)$, $K(1460)$, $\eta(1475)$, $K(1630)$, $\eta(1760)$, $\pi(1800)$ and $K(1830)$.
Pseudoscalar ground-state quarkonia have masses below 1 GeV -- thus there appear to exist sufficiently many candidates for the excited
$\bar{q}q$ pseudoscalars but a complication is that the $\eta(1405)$ and $\eta(1475)$ resonances have been claimed to represent a
single state denoted as $\eta(1440)$ \cite{eta1440}. In this article, $\eta(1295)$ is taken to represent the excited $\bar{n}n$ 
pseudoscalar isosinglet $\eta_N^E$ while the isotriplet state $\pi^E$ is assigned to $\pi(1300)$. Furthermore,
$\eta(1440)$ is assumed to represent the excited
pseudoscalar $\bar{s}s$ state $\eta_S^E$ due to the decay patterns (see Ref.\ \cite{eta1440}) and the mass difference
to $\eta(1295)$. The excited kaon $K^E$ is assigned to $K(1460)$.
 
\end{itemize}
$\,$\\
With these assignments, the following mass values are used to determine the unknown parameters in Eqs.\ (\ref{sigmaNE}) - (\ref{KE}):
(\textit{i}) $m_{\pi^E} = 1300 \pm 100$ MeV \cite{PDG}; (\textit{ii}) $m_{\eta_S^E} = 1432 \pm 10$ MeV \cite{eta1440};
(\textit{iii}) $m_{K^E} = 1460 \pm 73$ MeV \cite{PDG} and (\textit{iv}) $m_{\sigma_N^E} =  1790 \pm 35$ MeV \cite{BES} 
(the error here is average of the upper and lower values reported in Ref.\ \cite{BES}). Note that
no mass error for $K(1460)$ was obtained in analyses cited by the PDG; therefore, a generic
mass uncertainty of $\pm 5\%$ ($ \equiv \pm 73$ MeV) is used for this resonance in accordance with general precision aims of the model \cite{PGR,PKWGR}.
\\
\\
The above data allow for determination of the parameters as well as masses (including theoretical uncertainties stemming
from the experimental ones) in Eqs.\ (\ref{sigmaNE}) - (\ref{KE}). Results are
presented in Fig.\ \ref{Figure1} and compared to the experimental data.
\\
\\
Three masses are model predictions: for the scalar $\bar{s}s$ isosinglet $\sigma_S^E$,  
scalar kaon $ K_{0}^{\star E}$ and scalar isotriplet $a_0^E$.
They are (preliminarily): $m_{\sigma_S^E} = (1911 \pm 39)$ MeV, $m_{K_{0}^{\star E}} = (1851 \pm 37)$ MeV and 
$m_{a_0^E} = (1790 \pm 35)$ MeV. [Note that $m_{a_0^E} = m_{\sigma_N^E} \equiv m_{f_0(1790)}$ since the $a_0^E$-$\sigma_N^E$
mass splitting is determined by the large-$N_c$ suppressed parameter $\kappa_2$ that was set to zero in Eq.\ (\ref{sigmaNE}).]
There is currently no experimentally
known candidate for $a_0^E$; contrarily, the scalar $\bar{s}s$ isosinglet and the scalar kaon are very close to the (as yet uncofirmed
\cite{PDG}) $f_0(2020)$ and $K_0^\star (1950)$ resonances. It needs to be noted however that the $f_0$ channel in particular is
susceptible to effects of mixing that may alter mass patterns. Additionally, since the simulations of
pure-gauge Lattice QCD obtain the mass of the scalar glueball to be $\simeq 1.7$ GeV \cite{Morningstar}, we can conclude that our excited isosinglet $\bar{q}q$ mesons are sufficiently close to possibly influence glueball phenomenology. This is currently under investigation \cite{PG}. 

\section{Summary and Outlook}

Excited scalar and pseudoscalar $\bar{q}q$ mesons containing $u$, $d$ and $s$ quarks have been investigated in the Extended
Linear Sigma Model (eLSM) that, in addition to the excited states, also contains ground-state scalar, pseudoscalar, vector and
axial-vector degrees of freedom. The model implements the chiral and dilatational symmetries and their breaking as well the discrete
symmetries of QCD (charge conjugation, parity and time reversal). Assuming the resonances $\pi(1300)$ \cite{PDG},
$\eta(1440)$ \cite{eta1440}, $K(1460)$ \cite{PDG} and $f_0(1790)$ \cite{BES,LHCb} to represent excited $\bar{q}q$ states, masses of
further three excited mesons are predicted: $m_{\sigma_S^E} = (1911 \pm 39)$ MeV (scalar $\bar{s}s$ isosinglet), $m_{K_{0}^{\star E}} = (1851 \pm 37)$ MeV
(scalar kaon) and $m_{a_0^E} = (1790 \pm 35)$ MeV (scalar $\bar{n}n$ isotriplet). The first two masses are reasonably close to those
of the (as yet unconfirmed \cite{PDG}) $f_0(2020)$ and $K_0^\star (1950)$ resonances; there is currently no experimentally
known candidate for $a_0^E$. Additionally, the excited isosinglet $\bar{q}q$ mesons are sufficiently close in mass to the scalar glueball \cite{Morningstar} so as to possibly influence glueball phenomenology via mixing.
\\
\\
Further investigation of the excited mesons appears warranted. The importance of antiproton data has to be
particurly emphasised in this regard as $\bar{p}p$ collisions were a highly successful experimental tool in meson
physics of the past (see, e.g., Ref.\ \cite{Amsler:1995gf}) and consequently possess a high potential for discoveries in the
future, for example at PANDA@FAIR \cite{PANDA}.
\\
\\
{\bf Acknowledgments.}
We are grateful to F.~Br\"{u}nner, D.~Bugg, A.~Rebhan and D.~Rischke for collaboration and extensive
discussions. D.P. thanks Conference Organisers for their kind hospitality. 
This work is supported by the Austrian Science Fund FWF, project no.\ P26366.


\begin{thebibliography}{9}

\bibitem{AF} 
  D.~J.~Gross and F.~Wilczek,
  Phys.\ Rev.\ Lett.\  {\bf 30}, 1343 (1973);
  Phys.\ Rev.\ D {\bf 8}, 3633 (1973).
  H.~D.~Politzer,
  Phys.\ Rev.\ D {\bf 9}, 2174 (1974);
  Phys.\ Rept.\  {\bf 14}, 129 (1974).

  
\bibitem{Alkofer} 
  R.~Alkofer and J.~Greensite,
  J.\ Phys.\ G {\bf 34}, S3 (2007)
  [hep-ph/0610365].



\bibitem{PDG}
 K.~A.~Olive \textit{et al.} (Particle Data Group), Chin. Phys. C \textbf{38}, 090001 (2014) and 2015 update. 

\bibitem{Jaffe} 
  R.~L.~Jaffe and K.~Johnson,
  Phys.\ Lett.\ B {\bf 60}, 201 (1976);
  V.~A.~Novikov, M.~A.~Shifman, A.~I.~Vainshtein and V.~I.~Zakharov,
  Phys.\ Lett.\ B {\bf 86}, 347 (1979)
  [JETP Lett.\  {\bf 29}, 594 (1979)]
  [Pisma Zh.\ Eksp.\ Teor.\ Fiz.\  {\bf 29}, 649 (1979)];
  V.~A.~Novikov, M.~A.~Shifman, A.~I.~Vainshtein and V.~I.~Zakharov,
  Nucl.\ Phys.\ B {\bf 165}, 67 (1980).

  
  
\bibitem{LHCb} 
  R.~Aaij {\it et al.} [LHCb Collaboration],
  Phys.\ Rev.\ D {\bf 89}, no. 9, 092006 (2014)
  [arXiv:1402.6248 [hep-ex]].
  
  
\bibitem{BES} 
  M.~Ablikim {\it et al.} [BES Collaboration],
  Phys.\ Lett.\ B {\bf 607}, 243 (2005)
  [hep-ex/0411001].
 
  


\bibitem{glueballmixing} 
  C.~Amsler and F.~E.~Close,
  Phys.\ Rev.\ D {\bf 53}, 295 (1996)
  [hep-ph/9507326];
  W.~J.~Lee and D.~Weingarten,
  Phys.\ Rev.\ D {\bf 61}, 014015 (2000)
  [hep-lat/9910008];
  F.~E.~Close and A.~Kirk,
  Eur.\ Phys.\ J.\ C {\bf 21}, 531 (2001)
  [hep-ph/0103173];
  C.~Amsler and N.~A.~Tornqvist,
  Phys.\ Rept.\  {\bf 389}, 61 (2004);
  F.~E.~Close and Q.~Zhao,
  Phys.\ Rev.\ D {\bf 71}, 094022 (2005)
  [hep-ph/0504043];
  F.~Giacosa, T.~Gutsche, V.~E.~Lyubovitskij and A.~Faessler,
  Phys.\ Rev.\ D {\bf 72}, 094006 (2005)
  [hep-ph/0509247];
  M.~Albaladejo and J.~A.~Oller,
  Phys.\ Rev.\ Lett.\  {\bf 101}, 252002 (2008)
  [arXiv:0801.4929 [hep-ph]];
  V.~Mathieu, N.~Kochelev and V.~Vento,
  Int.\ J.\ Mod.\ Phys.\ E {\bf 18}, 1 (2009)
  [arXiv:0810.4453 [hep-ph]];
  P.~Zhang, L.~P.~Zou, J.~J.~Xie, J.~H.~Yoon and Y.~M.~Cho,
  arXiv:1606.02374 [hep-ph].

  
\bibitem{Janowski:2011gt} 
  S.~Janowski, D.~Parganlija, F.~Giacosa and D.~H.~Rischke,
  Phys.\ Rev.\ D {\bf 84}, 054007 (2011)
  [arXiv:1103.3238 [hep-ph]];
  S.~Janowski, F.~Giacosa and D.~H.~Rischke,
  Phys.\ Rev.\ D {\bf 90}, no. 11, 114005 (2014)
  [arXiv:1408.4921 [hep-ph]].
  
\bibitem{BPR} 
  F.~Br\"{u}nner, D.~Parganlija and A.~Rebhan,
  Phys.\ Rev.\ D {\bf 91}, no. 10, 106002 (2015)
  Erratum: [Phys.\ Rev.\ D {\bf 93}, no. 10, 109903 (2016)]
  [arXiv:1501.07906 [hep-ph]];
  F.~Br\"{u}nner and A.~Rebhan,
  Phys.\ Rev.\ Lett.\  {\bf 115}, no. 13, 131601 (2015)
  [arXiv:1504.05815 [hep-ph]];
  Phys.\ Rev.\ D {\bf 92}, no. 12, 121902 (2015)
  [arXiv:1510.07605 [hep-ph]].

\bibitem{PANDA} 
  M.~F.~M.~Lutz {\it et al.} [PANDA Collaboration],
  arXiv:0903.3905 [hep-ex].
  
\bibitem{NICA} 
  D.~Parganlija,
  arXiv:1601.05328 [hep-ph].
  
  
\bibitem{PG} 
  D.~Parganlija and F.~Giacosa, in preparation.

\bibitem{Freund:1968nf} 
  P.~G.~O.~Freund,
  Nuovo Cim.\ A {\bf 58}, 519 (1968);
  C.~T.~Chen-Tsai and T.~Y.~Lee,
  Phys.\ Rev.\ D {\bf 10}, 2960 (1974).
  
  
\bibitem{Ino:1984pi} 
  T.~Ino,
  Prog.\ Theor.\ Phys.\  {\bf 71}, 864 (1984);
  P.~Geiger,
  Phys.\ Rev.\ D {\bf 49}, 6003 (1994)
  [hep-ph/9311254];
  S.~M.~Fedorov and Y.~A.~Simonov,
  JETP Lett.\  {\bf 78}, 57 (2003)
  [Pisma Zh.\ Eksp.\ Teor.\ Fiz.\  {\bf 78}, 67 (2003)]
  [hep-ph/0306216];
  T.~Gutsche, V.~E.~Lyubovitskij and M.~C.~Tichy,
  Phys.\ Rev.\ D {\bf 79}, 014036 (2009)
  [arXiv:0811.0668 [hep-ph]];
  G.~Rupp, S.~Coito and E.~van Beveren,
  arXiv:1605.04260 [hep-ph].


\bibitem{Lacock:1996vy} 
  P.~Lacock {\it et al.} [UKQCD Collaboration],
  Phys.\ Rev.\ D {\bf 54}, 6997 (1996)
  [hep-lat/9605025];
  T.~Burch, C.~Gattringer, L.~Y.~Glozman, C.~Hagen, C.~B.~Lang and A.~Schafer,
  Phys.\ Rev.\ D {\bf 73}, 094505 (2006)
  [hep-lat/0601026];
  J.~J.~Dudek, R.~G.~Edwards, M.~J.~Peardon, D.~G.~Richards and C.~E.~Thomas,
  Phys.\ Rev.\ Lett.\  {\bf 103}, 262001 (2009)
  [arXiv:0909.0200 [hep-ph]];
  Phys.\ Rev.\ D {\bf 82}, 034508 (2010)
  [arXiv:1004.4930 [hep-ph]];
  J.~J.~Dudek {\it et al.} [Hadron Spectrum Collaboration],
  Phys.\ Rev.\ D {\bf 88}, no. 9, 094505 (2013)
  [arXiv:1309.2608 [hep-lat]].
  
  
\bibitem{Holl:2004fr} 
  A.~Holl, A.~Krassnigg and C.~D.~Roberts,
  Phys.\ Rev.\ C {\bf 70}, 042203 (2004)
  [nucl-th/0406030];
  A.~Holl, A.~Krassnigg, P.~Maris, C.~D.~Roberts and S.~V.~Wright,
  Phys.\ Rev.\ C {\bf 71}, 065204 (2005)
  [nucl-th/0503043];
  B.~L.~Li, L.~Chang, F.~Gao, C.~D.~Roberts, S.~M.~Schmidt and H.~S.~Zong,
  arXiv:1604.07415 [nucl-th].



\bibitem{Volkov:1997dd} 
  M.~K.~Volkov, D.~Ebert and M.~Nagy,
  Int.\ J.\ Mod.\ Phys.\ A {\bf 13}, 5443 (1998)
  [hep-ph/9705334];
  M.~K.~Volkov, V.~L.~Yudichev and D.~Ebert,
  J.\ Phys.\ G {\bf 25}, 2025 (1999)
  [JINR Rapid Commun.\  {\bf 6-92}, 5 (1998)]
  [hep-ph/9810470];
  M.~K.~Volkov and V.~L.~Yudichev,
  Int.\ J.\ Mod.\ Phys.\ A {\bf 14}, 4621 (1999)
  [hep-ph/9904226];
  Phys.\ Part.\ Nucl.\  {\bf 31}, 282 (2000)
  [Fiz.\ Elem.\ Chast.\ Atom.\ Yadra {\bf 31}, 576 (2000)]
  [hep-ph/9906371];
  Phys.\ Atom.\ Nucl.\  {\bf 63}, 1835 (2000)
  [Yad.\ Fiz.\  {\bf 63N10}, 1924 (2000)]
  [hep-ph/9905368];
  Phys.\ Atom.\ Nucl.\  {\bf 63}, 455 (2000)
  [Yad.\ Fiz.\  {\bf 63}, 527 (2000)];
  Eur.\ Phys.\ J.\ A {\bf 10}, 223 (2001)
  [hep-ph/0103003];
  Phys.\ Atom.\ Nucl.\  {\bf 65}, 1657 (2002)
  [Yad.\ Fiz.\  {\bf 65}, 1701 (2002)];
  A.~B.~Arbuzov, E.~A.~Kuraev and M.~K.~Volkov,
  Phys.\ Rev.\ C {\bf 82}, 068201 (2010)
  [arXiv:1007.1057 [hep-ph]];
  A.~V.~Vishneva and M.~K.~Volkov,
  Phys.\ Part.\ Nucl.\ Lett.\  {\bf 11}, 352 (2014)
  [arXiv:1312.1470 [hep-ph]];
  Int.\ J.\ Mod.\ Phys.\ A {\bf 29}, no. 24, 1450125 (2014)
  [arXiv:1403.1360 [hep-ph]].

\bibitem{Arndt:1999wx} 
  D.~Arndt and C.~R.~Ji,
  Phys.\ Rev.\ D {\bf 60}, 094020 (1999)
  [hep-ph/9905360].

  
\bibitem{Badalian:2002rc} 
  A.~M.~Badalian and B.~L.~G.~Bakker,
  Phys.\ Rev.\ D {\bf 66}, 034025 (2002)
  [hep-ph/0202246].

  
\bibitem{Nedelko:2016gdk} 
  S.~N.~Nedelko and V.~E.~Voronin,
  Phys.\ Rev.\ D {\bf 93}, no. 9, 094010 (2016)
  [arXiv:1603.01447 [hep-ph]].


\bibitem{Wagenbrunn:2006cs} 
  R.~F.~Wagenbrunn and L.~Y.~Glozman,
  Phys.\ Lett.\ B {\bf 643}, 98 (2006)
  [hep-ph/0605247];
  Phys.\ Rev.\ D {\bf 75}, 036007 (2007)
  [hep-ph/0701039].


\bibitem{PGR} 
  D.~Parganlija, F.~Giacosa and D.~H.~Rischke,
  Phys.\ Rev.\ D {\bf 82}, 054024 (2010)
  [arXiv:1003.4934 [hep-ph]].
\bibitem{PKWGR} 
  D.~Parganlija, P.~Kovacs, G.~Wolf, F.~Giacosa and D.~H.~Rischke,
  Phys.\ Rev.\ D {\bf 87}, no. 1, 014011 (2013)
  [arXiv:1208.0585 [hep-ph]].

\bibitem{Rosenzweig:1979ay} 
  C.~Rosenzweig, J.~Schechter and C.~G.~Trahern,
  Phys.\ Rev.\ D {\bf 21}, 3388 (1980).
  
  
\bibitem{Giacosa:2009qq} 
  F.~Giacosa,
  Eur.\ Phys.\ J.\ C {\bf 65}, 449 (2010)
  [arXiv:0907.3519 [hep-ph]].

  
 
\bibitem{eta1440} 
  J.~Z.~Bai {\it et al.} [BES Collaboration],
  Phys.\ Lett.\ B {\bf 440}, 217 (1998);
  J.~Z.~Bai {\it et al.} [BES Collaboration],
  Phys.\ Lett.\ B {\bf 476}, 25 (2000)
  [hep-ex/0002007].


\bibitem{Morningstar} 
  C.~J.~Morningstar and M.~J.~Peardon,
  Phys.\ Rev.\ D {\bf 60}, 034509 (1999)
  [hep-lat/9901004].


\bibitem{Amsler:1995gf} 
  C.~Amsler {\it et al.},
  Phys.\ Lett.\ B {\bf 342}, 433 (1995).


  
  
 
\end{thebibliography}
\end{document}